\documentclass[10pt,conference]{IEEEtran}
\usepackage{mdframed}
\usepackage{amsmath}
\usepackage{graphicx}
\usepackage{xspace}
\usepackage{hyperref}
\usepackage{xcolor}
\graphicspath{{Graphs/}}
\usepackage{caption}
\usepackage{subcaption}
\usepackage{pgfplots}
\pgfplotsset{compat=1.18}
\usetikzlibrary{patterns}
\usepackage[inline]{enumitem}
\usepackage{enumerate}
\usepackage{enumitem}
\usepackage{listings}
\usepackage{tabularx}
\usepackage{makecell}
\usepackage{tcolorbox}
\usepackage{float}
\usepackage{colortbl}
\usepackage{cite}

\tcbset{
  takeaway/.style={
    colback=gray!10,
    colframe=black,
    fonttitle=\bfseries,
    boxrule=0.5mm,
    arc=0mm,
    outer arc=0mm,
    width=\columnwidth, 
    boxsep=5pt,
    before skip=10pt,
    after skip=10pt,
    floatplacement=H, 
  }
}

\usepackage{orcidlink}
\hypersetup{
	plainpages=false,
	colorlinks,
	urlcolor=blue,
	linkcolor=blue,
	citecolor=red,
	bookmarksnumbered
}

\newcommand{\xl}[1]{\Xhline{#1\arrayrulewidth}}

\newcommand{\BfPara}[1]{{\noindent\bf#1.}\xspace}
\newcommand{\etal}{{\em et al.}\xspace}
\newcommand{\eg}{{\em e.g.},\xspace}
\newcommand{\ie}{{\em i.e.},\xspace}

\newcommand{\rrx}[1]{\textcolor{red}{#1}}
\newcommand{\ggx}[1]{\textcolor{green!60!black}{#1}}
\usepackage{pifont}
\usepackage[misc,geometry]{ifsym}

\newcommand{\ours}{{\sl xr-scope}\xspace}
\begin{document}

\title{Through the Looking Glass: LLM-Based Analysis of AR/VR Android Applications Privacy Policies}

\author{Abdulaziz Alghamdi \orcidlink{0009-0008-8508-9281} and David Mohaisen \orcidlink{0000-0003-3227-2505} \\
{\em University of Central Florida}}


\providecommand{\keywords}[1]
{	
  \textbf{\textit{Keywords---}} #1
}

\IEEEtitleabstractindextext{
\begin{abstract} This paper comprehensively analyzes privacy policies in AR/VR applications, leveraging BERT, a state-of-the-art text classification model, to evaluate the clarity and thoroughness of these policies. By comparing the privacy policies of AR/VR applications with those of free and premium websites, this study provides a broad perspective on the current state of privacy practices within the AR/VR industry. Our findings indicate that AR/VR applications generally offer a higher percentage of positive segments than free content but lower than premium websites. The analysis of highlighted segments and words revealed that AR/VR applications strategically emphasize critical privacy practices and key terms. This enhances privacy policies' clarity and effectiveness. \\
\keywords{AR, VR, Privacy Policy, LLM, Machine Learning}
\end{abstract}}

\maketitle
\IEEEdisplaynontitleabstractindextext
\IEEEpeerreviewmaketitle

\section{Introduction}

Privacy policies inform users how their data is collected, used, and shared by applications \cite{ZimmeckB14, AdamsBBMPR18}. Despite their importance, these policies are often long, complex, and difficult to comprehend \cite{LehmanAKLT22}, leading to potential misalignments between stated and actual data practices, which may result in privacy violations \cite{SlavinWHHKBBN16a}. Tools like PVDetector \cite{SlavinWHHKBBN16a} have been developed to automatically analyze policies and detect such misalignments, helping to prevent legal issues \cite{AhmadCTC20}.

This issue is especially relevant for AR/VR applications, which gather a broad range of sensitive data, including biometrics and environmental details \cite{LimYHK22, MaragkoudakiK22}. As AR/VR technologies grow, ensuring their privacy policies are transparent and compliant with regulations is crucial for user trust \cite{HarborthHTR19, HarborthP21}. However, these policies are often overly complex and lengthy, discouraging user engagement \cite{LiuWSZS18, LimYHK22}.

Given the sensitive nature of AR/VR apps used across various industries, robust privacy practices and clear communication of these policies are essential. Despite this, research on AR/VR privacy practices remains limited \cite{LimYHK22}. This study aims to address this gap by using BERT to evaluate AR/VR privacy policies' transparency and comparing them with those of free and premium websites, identifying areas for improvement.

\BfPara{Contribution} \begin{enumerate*}
\item We compiled and analyzed a dataset of privacy policies from AR/VR applications. 
\item We utilized BERT, a state-of-the-art text classification model, to evaluate the clarity and comprehensiveness of these privacy policies, highlighting current strengths and weaknesses.
\item We compared AR/VR applications' privacy policies with free and premium websites, offering a broader context for understanding AR/VR privacy practices' relative transparency.
\end{enumerate*}

\BfPara{Organization} We review the related work in~\textsection\ref{sec:related}, methods in \textsection\ref{sec:pipeline}, results and discussion in \textsection\ref{sec:results} and conclusion in \textsection\ref{sec:conclusion}.

\section{Related Work}\label{sec:related}

\begin{table*}[t]
\centering
\caption{An overview of related research on privacy policy analysis, NLP techniques, and AR/VR applications.}
\label{table:overview}
\scalebox{0.99}{
\begin{tabular}{ lcccccl }
\xl{2}
{\bf Author} & {\bf Year} & {\bf Focus} & {\bf Dataset Size} & {\bf Method} & {\bf Outcome} \\
 \xl{1}
Devlin \etal~\cite{DevlinCLT19} & 2019 & NLP Model & N/A & BERT & Language understanding \\
Alabduljabbar \etal~\cite{AlabduljabbarM22} & 2022 & Privacy Policy Analysis & 720 policies & NLP, BERT & Policy categorization \\
Guo \etal~\cite{GuoDLZXH24} & 2024 & VR Analysis & 500 VR apps & VR-SP & Security and privacy detection \\
Elluri \etal~\cite{ElluriCJFJ21} & 2021 & Policy Compliance & 3,000 policies & BiLSTM, BERT & Compliance analysis with GDPR \\
Wilson \etal~\cite{WilsonSDLCLAZSRNHRS16} & 2016 & Privacy Policies & 115 policies & Text Mining & Policy extraction \\
Andow \etal~\cite{AndowMWWERSX19} & 2019 & Policy Tools & 11,430 policies & PolicyLint & Contradictions in policies \\
Yu \etal~\cite{YuLLZ16} & 2016 & Policy Trustworthiness & 1,197 apps & PPChecker & Identified issues in privacy policies \\
\xl{1}
\ours & 2024 & AR/VR Policy Analysis & 302 policies & NLP, BERT & Privacy practices evaluation \\
\xl{2}
\end{tabular}}
\end{table*}

Table \ref{table:overview} reviews some of the prior work on privacy policy analysis, NLP (\ie Natural Language Processing) techniques, and AR/VR applications to provide a context for \ours work. Most of the studies explored in this section are related to privacy policy analysis using machine learning models, and we exclude other studied focused on different aspects, although they are plentiful and deserve a separate study~\cite{DBLP:conf/csonet/AlasmaryA0CNM18,DBLP:data/10/KinoonAAM24,DBLP:conf/icdcs/AbusnainaKA0AM19,DBLP:journals/corr/abs-1902-04416,DBLP:journals/iotj/AlasmaryAAAAWNA22,DBLP:conf/raid/AbusnainaAAAJNM22}. These studies focus on BERT in particular and its applications in AR/VR. Based on Table \ref{table:overview}, no previous study has comprehensively analyzed AR/VR privacy policies using BERT. The existing works focus on various aspects such as privacy policy extraction, data protection in AR/VR, and the use of NLP models for text classification.

\BfPara{Privacy Policy Analysis} Privacy policies inform users about data collection, use, and protection practices~\cite{AlabduljabbarAMY21,AlabduljabbarM22,AlabduljabbarAMM21,AlghamdiAAM24}. Numerous studies have focused on analyzing these policies. For instance, Alabduljabbar \etal~\cite{AlabduljabbarM22} conducted a comprehensive analysis of privacy policies using a BERT-based technique, categorizing segments into predefined categories and showing trends in the analyzed policies for the presence or absence of various collection, use, and protection. Wilson \etal~\cite{WilsonSDLCLAZSRNHRS16} and Andow \etal~\cite{AndowMWWERSX19} developed tools for extracting and analyzing privacy policies to identify potential misalignment between the stated and the actual practices. These studies emphasize the need for advanced NLP techniques to help better understand and classify privacy policies. However, this analysis is done mostly for websites and does not consider privacy policies in AR/VR apps. Yu \etal~\cite{YuLLZ16} introduced PPChecker, a system that uses NLP and program analysis to identify issues in privacy policies: incompleteness, incorrectness, and inconsistency, which is an issue prevalent in other related applications~\cite{DBLP:conf/dimva/MohaisenA14,DBLP:conf/wisa/AlthebeitiM23,DBLP:conf/wise/AlkinoonAAMSM24,DBLP:conf/ccs/AlthebeitiFCM23,DBLP:conf/securecomm/AnwarKNM18,DBLP:journals/tdsc/AnwarACLM22}.

\BfPara{NLP Techniques in Privacy Analysis} NLP techniques, e.g., BERT, have advanced text classification and analysis that are used for a range of tasks for security applications~\cite{MohaisenA14a,ShenVMKZ17,WestM14,MohaisenA13}. Devlin \etal~\cite{DevlinCLT19} introduced BERT, a model that has revolutionized NLP by providing a deep contextual understanding of text. This model has been utilized in various domains, including privacy policy analysis, due to its ability to capture nuanced meanings and relationships within text. Elluri \etal~\cite{ElluriCJFJ21} demonstrated BERT's effectiveness in enhancing text analysis and classification and showed significant improvements in accuracy and robustness, making BERT an ideal choice for analyzing complex privacy policy documents.

\BfPara{AR/VR Privacy Policies} 
Although studies by Harborth \etal~\cite{HarborthHTR19} and Lim \etal~\cite{LimYHK22} highlight the critical importance of transparent and comprehensive privacy policies in this domain, quantifying the gap in the practice from that of the expected privacy policy structure and compliance is still lacking.



\section{Privacy Policy Analysis Pipeline}\label{sec:pipeline}

Our methodology systematically collects and processes privacy policies from AR/VR applications, employing advanced NLP techniques. BERT was tuned using an annotated dataset containing segments of privacy policies, which had been meticulously curated. This fine-tuning ensured that the analysis maintained alignment with prior research and achieved a semantically robust mapping of policies to high-level attributes.

This assessment within AR/VR applications was conducted by categorizing and evaluating positive segments in which specific policy elements can be identified and highlighting segments and key terms. Furthermore, this approach enabled us to assess the richness and expressiveness of these privacy policies compared to those of other domains, such as website privacy policies. \ours analysis provides critical insights into the nuances of privacy policy articulation in AR/VR applications. This contributes to the broader understanding of privacy practices in emerging technological contexts.

\begin{figure}[t]
\centering
\includegraphics[scale=0.14]{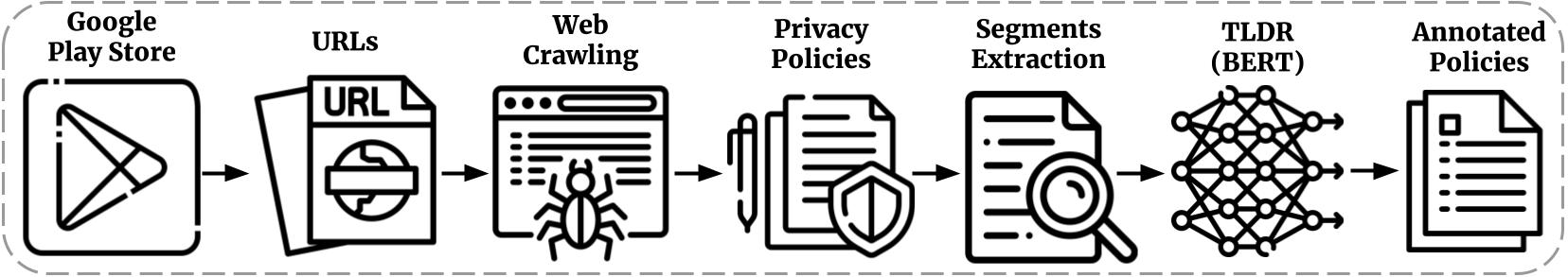}
\caption{Privacy Policy Pipeline}
\label{fig: Pipeline}
\end{figure}

\subsection{Dataset Scraping \& Transformation}

Each app's privacy policy was extracted using advanced web scraping techniques. Starting with 408 URLs from \cite{AlghamdiAAM24}, the URLs were filtered for relevance and accessibility. This automated method reduced manual effort and improved accuracy, compiling a comprehensive list of privacy policy URLs efficiently. HTML content from the policies was retrieved using the URLs, addressing edge cases such as non-English content, image-based policies, inaccessible documents, and apps no longer available on Google Play. Robust error handling ensured data integrity, yielding 302 usable privacy policies. These were obtained by systematically reading URLs from structured files and retrieving HTML content, resulting in 302 policies in text or HTML format, as shown in Table~\ref{tab:availability}. Advanced parsing techniques were applied to extract relevant text from the HTML files, removing extraneous elements for clean, usable text. This pre-processing prepared the data for analysis, focusing on the substantive content. The extraction and transformation process is visualized in Figure \ref{fig: Pipeline}.

\begin{table}[t]
\centering
\caption{Availability of privacy policies in \ours.}
\label{tab:availability}
\scalebox{0.99}{
\begin{tabular}{lcc}
\xl{2}
\textbf{Type} & {\bf Included?} & \textbf{Applications}\\ 
\xl{1}
Image & \textcolor{red}{\ding{54}} & 1 \\
No link or document &\textcolor{red}{\ding{54}} & 16\\
Not in English &\textcolor{red}{\ding{54}} & 47 \\
Apps not available &\textcolor{red}{\ding{54}} & 42 \\
Text &\textcolor{green!50!black}{\ding{52}} & 22\\
HTML &\textcolor{green!50!black}{\ding{52}} & 280\\
\xl{2}
\end{tabular}}
\end{table}

\subsection{Dataset Processing}

The extracted text from privacy policies was tokenized into words, with non-alphabetic tokens and stopwords removed to focus on key terms like "information," "personal," "data," "policy," "service," and "privacy." To ensure dataset quality, we excluded files without the term `privacy' and those with fewer than two instances of key terms such as `information,' `personal,' `data,' and `policy.' This process used saved word counts to eliminate irrelevant files, refining the dataset to only substantive policies—further refinement involved excluding short, non-informative paragraphs (\ie fewer than five words). The cleaned text was saved to enhance data for robust analysis. 

All paragraphs were annotated by converting them to lowercase. The total number of paragraphs and words across all documents was aggregated, with average metrics per document calculated to assess the structure and verbosity of privacy policies. For comparative analysis, different groups (\eg categories or types of applications) were examined. The number of paragraphs and words for each group was calculated using the annotated data, with averages computed for comparison. This comparison helped identify trends and differences in privacy policy disclosures among various application types, improving the understanding of privacy practices in different contexts.

\subsection{Privacy Policy Categories}

The methodology from \cite{AlabduljabbarM22, AlabduljabbarAMY21} was used to categorize privacy policy segments into nine categories: First Party Use, Third Party Sharing, User Choice, User Access, Data Retention, Data Security, Policy Change, Do Not Track, and Specific Audiences. Initially applied to free content and premium websites, this framework is relevant for analyzing AR/VR apps' privacy policies, aiming to compare their privacy practices with \cite{AlabduljabbarM22}.  First Party Use addresses how app developers use collected data internally, while Third Party Sharing focuses on data shared with external entities. User Choice covers users' data control options, and User Access outlines how users manage their data. Data Retention addresses how long data is stored, and Data Security focuses on data protection measures. Policy Change explains user notification of privacy updates, and Do Not Track assesses the app's response to tracking preferences. Specific Audiences focuses on handling data for vulnerable groups like children \cite{WilsonSDLCLAZSRNHRS16}, highlighting the app's approach to privacy for different demographics.

\subsection{BERT Analysis}

BERT (Bidirectional Encoder Representations from Transformers) is a language model developed by Google \cite{DevlinCLT19}, excelling in various NLP tasks due to its bidirectional context understanding \cite{ElluriCJFJ21, AhmadCLNTC21}. BERT was chosen for analyzing privacy policies of AR/VR apps because of its strong performance in text classification tasks that require deep contextual comprehension. The model was fine-tuned using data from \cite{AlabduljabbarM22}, following their methodology to enable comparison of results. 

The training input consisted of category labels for each paragraph, annotated segments from the OPP-115 dataset, and document lengths, ensuring consistency with \cite{AlabduljabbarM22}'s approach. Hyperparameters were optimized, and data augmentation was employed to improve robustness. The dataset was split into training and testing sets with balanced categories, and cross-validation was performed to assess generalizability.

Training steps included loading data, splitting it into training/testing sets, BERT-specific pre-processing, model training, and evaluation on the test data. BERT classified privacy policy segments into predefined categories from \cite{AlabduljabbarM22}, allowing for automated analysis of privacy practices.

Positive segments were identified as clear explanations of data handling practices, essential for evaluating the transparency and comprehensiveness of privacy policies. The percentage of positive segments per category was calculated to assess how well the policies communicated privacy practices.

Highlighted segments and words in the policies were analyzed to identify key practices. This analysis calculated the percentage of significant segments/words within each category, focusing on critical data collection, sharing, retention, and security information. This highlighted crucial aspects of the policies and provided metrics for evaluating their effectiveness in conveying essential privacy information to users.


\section{Results \& Discussion}\label{sec:results}

In this section, the results of the analysis of AR/VR application privacy policies are presented, leveraging BERT for classification and evaluation. Various aspects are examined, including word and paragraph counts, BERT training accuracy, positive segments, highlighted segments, and highlighted words. By comparing these findings with the results from \cite{AlabduljabbarM22}, insights into the transparency and comprehensiveness of privacy policies in the AR/VR domain are provided. This analysis highlights the current state of privacy practices in AR/VR applications and identifies areas for improvement.

\subsection{Words and Paragraphs Count}

\BfPara{Overall}
The analysis began with a comprehensive examination of the total word and paragraph counts in AR/VR application privacy policies. A total of 240 policies were processed, resulting in a dataset containing 25,135 paragraphs and 930,225 words. This yields an average of 104.73 paragraphs and 3,875.94 words per policy. These metrics indicate that privacy policies for AR/VR applications tend to be relatively detailed. This is consistent with the need for thorough explanations of data practices in technology-intensive domains.

Comparatively, the average length of AR/VR privacy policies, in terms of both paragraphs and words, aligns more closely with the detailed policies of premium websites analyzed by \cite{AlabduljabbarM22}, which often feature comprehensive privacy disclosures. This suggests that AR/VR applications, similar to premium websites, prioritize detailed privacy statements to address unique data practices.

\begin{table}[t]
\centering
\caption{The statistics of \ours's dataset.}
\label{tab:overall_words_paragraphs}
\scalebox{0.99}{
\begin{tabular}{ lc }
 \xl{2}
{\bf Metric} & {\bf Count} \\ 
\xl{1}
Total Policies & 240 \\
Total Paragraphs & 25,135 \\
Avg. Paragraphs & 104.73 \\
Total Words & 930,225 \\
Avg. Words & 3,875.94 \\
\xl{2}
\end{tabular}}
\end{table}

\BfPara{Groups}
To gain a deeper understanding of the variations in privacy policy content across different types of AR/VR applications, the policies were categorized into specific groups based on their primary functions. These groups included education, games, entertainment, simulation, tools, video players, business, art \& photo, casual, books \& news, social \& communication, lifestyle, productivity, sports, health, travel \& maps, and shopping.

\ours findings revealed significant variations among these groups. For instance, the Entertainment group exhibited the highest average number of paragraphs per policy at 260.6, indicating particularly detailed and extensive privacy policies. In contrast, the Travel \& Maps group had the lowest average at 25 paragraphs, suggesting shorter privacy policies. Similarly, in terms of word counts, the Entertainment group had the highest total word count (297,159), reflecting its high average paragraph count, while Travel \& Maps had the lowest word count (393), aligning with its fewer paragraphs.

These disparities highlight how different AR/VR application categories prioritize privacy disclosures. Applications in categories such as entertainment and games, which may involve more complex data interactions, tend to provide more detailed privacy policies. Conversely, categories like travel and maps might have simpler data practices, resulting in shorter policies.

\begin{table}[ht!]
\centering
\caption{Group-based statistics of \ours's dataset.}
\label{tab:group_words_paragraphs}
\scalebox{0.93}{\begin{tabular}{lrrrrr}
\xl{2}
{\bf Group} & {\bf Policies} & {\bf Para.} & {\boldmath$\mu$ \bf Para.} & {\bf Words} & {\boldmath$\mu$ \bf Words} \\
\xl{1}
Education & 48 & 3,172 & 66.08 & 122,376 & 2,549.50 \\
Games & 30 & 7,873 & 262.43 & 221,499 & 7,383.31 \\
Entertainment & 30 & 7,818 & 260.60 & 297,159 & 9,905.30 \\
Simulation & 28 & 1,441 & 51.46 & 56,676 & 2,024.86 \\
Tools & 21 & 1,755 & 83.57 & 68,557 & 3,264.62 \\
Video Players & 18 & 484 & 26.89 & 21,960 & 1,220.00 \\
Business & 18 & 163 & 9.06 & 7,406 & 411.44 \\
Art \& Photo & 18 & 140 & 7.78 & 9,716 & 539.78 \\
Casual & 18 & 450 & 25.00 & 15,678 & 871.00 \\
Books \& News & 18 & 488 & 27.11 & 18,536 & 1,029.78 \\
Social \& Comm. & 16 & 359 & 22.44 & 14,527 & 908.94 \\
Lifestyle & 15 & 153 & 10.20 & 6,915 & 461.00 \\
Productivity & 15 & 422 & 28.13 & 15,774 & 1,051.60 \\
Sports & 10 & 175 & 17.50 & 7,455 & 745.50 \\
Health & 10 & 141 & 14.10 & 6,750 & 675.00 \\
Travel \& Maps & 1 & 25 & 25.00 & 393 & 393.00 \\
Shopping & 5 & 76 & 15.20 & 3,417 & 683.40 \\
\xl{2}
\end{tabular}}
\end{table}


\subsection{BERT Training}

\BfPara{Evaluation Metrics}
The BERT model was evaluated using multiple metrics to ensure a comprehensive assessment of its performance. Accuracy was determined as the proportion of correctly classified segments out of the total number of segments. Precision and recall were calculated for each category. Precision indicates the proportion of true positive predictions among all positive predictions, and recall reflects the proportion of true positive predictions among all actual positives. The F1-score, which is the harmonic mean of precision and recall, was utilized to provide a balanced measure of the model's performance. The overall performance of the model was assessed by averaging these metrics across all categories, offering a holistic view of its classification capabilities.

\begin{table}[htbp]
\centering
\caption{Comparison of the accuracy with different works.}
\label{tab:accuracy_comparison}
\small
\scalebox{0.93}{\begin{tabular}{@{}l@{\hspace{1mm}}c@{\hspace{1mm}}c@{\hspace{1mm}}c@{\hspace{1mm}}c@{\hspace{1mm}}c@{\hspace{1mm}}}
\xl{2}
{\bf Category} & {\bf \ours} & {\bf TLDR\cite{AlabduljabbarM22}} & {\bf Wilson\cite{WilsonSDLCLAZSRNHRS16}} & {\bf Harkous\cite{HarkousFLSSA18}} & {\bf Liu\cite{LiuWSZS18}} \\ 
\xl{1}
1st Party & 0.93 & 0.94 & 0.75 & 0.79 & 0.81 \\
3rd Party & 0.93 & 0.89 & 0.70 & 0.79 & 0.79 \\
User Choice & 0.96 & 0.85 & 0.61 & 0.74 & 0.70 \\
User Access & 0.98 & 0.91 & 0.61 & 0.80 & 0.82 \\
Data Retention & 0.99 & 0.87 & 0.16 & 0.71 & 0.43 \\
Data Security & 0.98 & 0.88 & 0.67 & 0.85 & 0.80 \\
Policy Change & 0.99 & 0.95 & 0.75 & 0.88 & 0.85 \\
Do Not Track & 1.00 & 1.00 & 1.00 & 0.95 & 1.00 \\
Audiences & 0.98 & 0.94 & 0.70 & 0.95 & 0.85 \\
\xl{1}
Overall & 0.97 & 0.91 & 0.66 & 0.83 & 0.78 \\
\xl{2}
\end{tabular}}
\end{table}

\BfPara{Training Process and Results}
The BERT model was initialized by downloading the pretrained BERT model (\textit{uncased\_L-12\_H-768\_A-12.zip}) and extracting it for use. The training data was preprocessed with a maximum sequence length of 512 tokens, and the model was fine-tuned using the onecycle policy with a maximum learning rate of 2e-05.

During training, the model's performance improved significantly over the epochs across all categories. For instance, in the 1st Party Use category, the model's accuracy increased from 80.5\% in the first epoch to 99.68\% in the final epoch. Similarly, the Do Not Track category demonstrated remarkable performance with an accuracy of 100\% achieved in several epochs. The detailed training results for all categories are summarized below (initial accuracy in epoch 1 vs. epoch 10):
\begin{enumerate*}
    \item {\em 1st Party Use}: 80.5\% vs. 99.68\%.
    \item {\em 3rd Party Sharing}: 78.17\% vs. 99.84\%.
    \item {\em User Choice}: 91.53\% vs. 99.87\%.
    \item {\em User Access}: 96.79\% vs. 99.84\%.
    \item {\em Data Retention}: 95.86\% vs. 99.91\%.
    \item {\em Data Security}: 93.65\% vs. 99.90\%.
    \item {\em Policy Change}: 97.16\% vs. 99.87\%.
    \item {\em Do Not Track}: 99.24\% vs. 100\%.
    \item {\em Specific Audiences}: 95.26\% vs. 99.87\%.
\end{enumerate*}

\BfPara{Reasons for Higher Results}
Despite using the same input data and similar training code as \cite{AlabduljabbarM22}, \ours BERT model achieved higher performance metrics. Several factors could contribute to this difference, including the hyperparameter tuning, the data augmentation and preprocessing, model initialization, training environment, and regularization technique.

\subsection{Positive Segments}

In this section, positive segments identified in AR/VR privacy policies are analyzed. Positive segments are those paragraphs that clearly articulate privacy practices and policies in a positive light, providing transparency and reassurance to users. The analysis results are compared with those from \cite{AlabduljabbarM22}, examining both overall trends and specific group differences.

\BfPara{Overall Comparison}
Table \ref{tab:positive_segments_overall} presents the overall percentage of positive segments across all categories. The results indicate that AR/VR applications have a higher percentage of positive segments than free content but a lower percentage than premium websites. This reflects a greater emphasis on transparency and user trust in premium websites compared to AR/VR applications.


\begin{table}[ht!]
\centering
\caption{Comparison of the distribution of the positive segments across various categories. $\Delta_F$ captures the difference between the distribution in \ours and the other group.}
\label{tab:positive_segments_overall}
\scalebox{0.99}{\begin{tabular}{ lccccc }
 \xl{2}
{\bf Category} & {\bf \ours} & {\bf Free} & {\bf $\Delta_F$} & {\bf Premium} & {\bf $\Delta_P$} \\ 
\xl{1}
1st Party Use & 97.08 & 86.90 & \ggx{+10.18} & 95.73 & \ggx{+1.35} \\
3rd Party Sharing & 92.50 & 84.52 & \ggx{+7.98} & 89.69 & \ggx{+2.81} \\
User Choice & 57.92 & 52.38 & \ggx{+5.54} & 79.27 & \rrx{-21.35} \\
User Access & 39.58 & 50.00 & \rrx{-10.42} & 65.81 & \rrx{-26.23} \\
Data Retention & 42.08 & 30.95 & \ggx{+11.13} & 57.26 & \rrx{-15.18} \\
Data Security & 75.00 & 67.86 & \ggx{+7.14} & 75.00 & 0.00 \\
Policy Change & 83.33 & 71.43 & \ggx{+11.90} & 72.22 & \ggx{+11.11} \\
Do Not Track & 14.17 & 12.70 & \ggx{+1.47} & 21.58 & \rrx{-7.41} \\
Specific Audiences & 77.50 & 67.86 & \ggx{+9.64} & 74.15 & \ggx{+3.35} \\
\xl{1}
Average & 64.35 & 58.29 & \ggx{+6.06} & 70.08 & \rrx{-5.73} \\
\xl{2}
\end{tabular}}
\end{table}

The comparison shows that AR/VR applications have a higher percentage of positive segments than free content but a lower percentage than premium websites. This suggests that while AR/VR applications are more transparent and provide better privacy assurances than free content, they still lag behind premium websites regarding overall positive segments. Categories like 1st Party Use, 3rd Party Sharing, Data Retention, and Policy Change exhibit significant positive differences between AR/VR apps and both free and premium websites.

The average value presented in Table \ref{tab:positive_segments_overall} is the mean value of all categories. It is calculated by summing the percentages of all categories and dividing by the total number of categories. This average provides a general overview of how positive segments are distributed across different privacy policy categories.

\BfPara{Groups Comparison}
In addition to the overall comparison, the positive segments for different groups of AR/VR applications were also analyzed. Table \ref{tab:positive_segments_groups} shows the percentage of positive segments for each group.

\begin{table}[ht!]
\centering
\caption{Comparison of the mean distribution value (\boldmath$\mu$) of the positive segments in \ours groups and other groups.}
\label{tab:positive_segments_groups}
\scalebox{0.99}{
\begin{tabular}{ lccc }
 \xl{2}
{\bf Category} & {\boldmath$\mu$ \bf \ours} & {\boldmath$\mu$ \bf Free} & {\boldmath$\mu$ \bf Premium} \\ 
\xl{1}
1st Party Use & 98.28 & 89.69 & 94.64 \\
3rd Party Sharing & 85.05 & 86.69 & 89.00 \\
User Choice & 64.86 & 59.06 & 79.71 \\
User Access & 38.82 & 44.33 & 67.20 \\
Data Retention & 38.73 & 34.92 & 59.61 \\
Data Security & 69.58 & 67.97 & 75.78 \\
Policy Change & 75.19 & 72.41 & 71.75 \\
Do Not Track & 8.82 & 12.25 & 22.87 \\
Specific Audiences & 63.72 & 68.97 & 75.68 \\
\xl{2}
\end{tabular}}
\end{table}

The group comparison reveals that AR/VR applications generally perform better or on par with free websites in terms of positive segments. However, there are some categories, such as Do Not Track and Specific Audiences, where AR/VR apps have lower percentages than other domains.

\BfPara{Observations and Explanations}
The high percentage of positive segments in AR/VR applications can be attributed to several factors: 
\begin{enumerate*}
    \item \textbf{Increased Focus on Transparency}: AR/VR applications often handle more sensitive data and have a higher level of user interaction, necessitating an increased focus on transparency to build user trust.
    \item \textbf{Regulatory Compliance}: Stricter privacy regulations and guidelines for AR/VR technologies may encourage developers to provide more comprehensive and clear privacy policies.
    \item \textbf{User Expectations}: Users of AR/VR applications may have higher expectations regarding privacy, prompting developers to be more transparent about their data practices.
\end{enumerate*}

Overall, \ours analysis indicates that AR/VR applications are making significant strides in privacy transparency. This is evidenced by the higher percentage of positive segments than free content and the targeted emphasis on critical privacy practices. However, they still have room to improve compared to premium websites.


\subsection{Highlighted Segments}

This section focuses on analyzing highlighted segments within privacy policies for AR/VR applications. Highlighted segments refer to those parts of the privacy policy that explicitly emphasize key privacy practices. This is often through bold text, headings, or other visual markers. The results are compared with those from \cite{AlabduljabbarM22} to evaluate the prominence and clarity of privacy practices in AR/VR applications.

\BfPara{Overall Comparison}
Table \ref{tab:highlighted_segments_overall} presents the overall percentage of highlighted segments across all categories. The results suggest that AR/VR applications have a varied distribution of highlighted segments compared to free and premium websites.

\begin{table}[ht!]
\centering
\caption{Comparison of the distribution of the highlighted segments in \ours and other groups.}
\label{tab:highlighted_segments_overall}
\scalebox{0.99}{
\begin{tabular}{ lccccc }
 \xl{2}
{\bf Category} & {\bf \ours} & {\bf Free} & {\bf $\Delta_F$} & {\bf Premium} & {\bf $\Delta_P$} \\
\xl{1}
1st Party Use & 35.51 & 25.76 & \ggx{+9.75} & 32.91 & \ggx{+2.60} \\
3rd Party Sharing & 19.06 & 16.00 & \ggx{+3.06} & 15.77 & \ggx{+3.29} \\
User Choice & 4.60 & 5.70 & \rrx{-1.10} & 6.12 & \rrx{-1.52} \\
User Access & 1.45 & 3.23 & \rrx{-1.78} & 3.14 & \rrx{-1.69} \\
Data Retention & 2.12 & 2.43 & \rrx{-0.31} & 1.89 & \ggx{+0.23} \\
Data Security & 3.54 & 3.39 & \ggx{+0.15} & 2.62 & \ggx{+0.92} \\
Policy Change & 2.57 & 2.22 & \ggx{+0.35} & 2.65 & \rrx{-0.08} \\
Do Not Track & 0.21 & 0.45 & \rrx{-0.24} & 0.31 & \rrx{-0.10} \\
Specific Audiences & 4.08 & 7.34 & \rrx{-3.26} & 8.37 & \rrx{-4.29} \\
\xl{1}
Overall & 62.29 & 58.96 & \ggx{+3.33} & 64.33 & \rrx{-2.04}\\
\xl{2}
\end{tabular}}
\end{table}

The overall comparison reveals that AR/VR applications generally perform similarly to or slightly better than free websites in terms of highlighted segments. However, they still lag behind premium websites in some categories. Notable categories with higher percentages include 1st Party Use and 3rd Party Sharing, indicating that these areas are more prominently emphasized in AR/VR applications.

\BfPara{Groups Comparison}
In addition to the overall comparison, the highlighted segments for different groups of AR/VR applications were also examined. Table \ref{tab:highlighted_segments_groups} shows the percentage of highlighted segments for each group.

\begin{table}[ht!]
\centering
\caption{Comparison of the mean distribution value (\boldmath$\mu$) of the highlighted segments in \ours groups, free websites groups, and premium websites groups privacy policies.}
\label{tab:highlighted_segments_groups}
\scalebox{0.99}{
\begin{tabular}{ lccc }
\xl{2}
{\bf Category} & {\boldmath$\mu$ \bf \ours} & {\boldmath$\mu$ \bf Free} & {\boldmath$\mu$ \bf Premium} \\
\xl{1}
1st Party Use & 98.28 & 89.69 & 94.64 \\
3rd Party Sharing & 85.05 & 86.69 & 89.00 \\
User Choice & 64.86 & 59.06 & 79.71 \\
User Access & 38.82 & 44.33 & 67.20 \\
Data Retention & 38.73 & 34.92 & 59.61 \\
Data Security & 69.58 & 67.97 & 75.78 \\
Policy Change & 75.19 & 72.41 & 71.75 \\
Do Not Track & 8.82 & 12.25 & 22.87 \\
Specific Audiences & 63.72 & 68.97 & 75.68 \\
\xl{2}
\end{tabular}}
\end{table}

The group comparison indicates that AR/VR applications have varying levels of highlighted segments across different categories. While categories like 1st Party Use and 3rd Party Sharing show higher emphasis, User Choice and User Access reveal lower percentages than free and premium websites.

\BfPara{Observations and Explanations}
The distribution of highlighted segments in AR/VR applications suggests a targeted approach to emphasizing specific privacy practices. Possible reasons include:
\begin{enumerate*}
    \item \textbf{Targeted Emphasis}: Developers might prioritize highlighting critical privacy practices relevant to AR/VR users.
    \item \textbf{Regulatory Focus}: Emphasis on certain categories could be driven by regulatory requirements specific to AR/VR technologies.
    \item \textbf{User Experience}: Enhanced user experience in AR/VR applications may lead developers to emphasize key privacy segments for better comprehension.
\end{enumerate*}

Overall, AR/VR applications exhibit a strategic approach to privacy segments, reflecting a comprehensive understanding of user needs and regulatory demands.


\subsection{Highlighted Words}

\BfPara{Introduction}
In this section, highlighted words in AR/VR privacy policies are evaluated. Highlighted words are those that are frequently used and emphasized in the context of privacy, such as `data,' `personal,' and `information.' The frequency and emphasis of these words are compared with the results from \cite{AlabduljabbarM22} to assess privacy communications focus areas.

\BfPara{Overall Comparison}
Table \ref{tab:highlighted_words_overall} presents the overall percentage of highlighted words across all categories. The results highlight the emphasis on certain key terms in AR/VR applications compared to free and premium websites.

\begin{table}[ht!]
\centering
\caption{Comparison of the distribution of the highlighted words in \ours and other groups. All numbers are \%.}
\label{tab:highlighted_words_overall}
\scalebox{0.99}{
\begin{tabular}{ lccccc }
 \xl{2}
{\bf Category} & {\bf \ours} & {\bf Free} & {\bf $\Delta_F$} & {\bf Premium} & {\bf $\Delta_P$} \\
\xl{1}
1st Party Use & 36.05 & 40.45 & \rrx{-4.40} & 31.41 & \ggx{+4.64} \\
3rd Party Sharing & 28.46 & 19.15 & \ggx{+9.31} & 21.04 & \ggx{+7.42} \\
User Choice & 6.04 & 6.29 & \rrx{-0.25} & 6.42 & \rrx{-0.38} \\
User Access & 1.86 & 3.56 & \rrx{-1.70} & 3.96 & \rrx{-2.10} \\
Data Retention & 1.97 & 2.39 & \rrx{-0.42} & 3.40 & \rrx{-1.43} \\
Data Security & 5.43 & 2.72 & \ggx{+2.71} & 4.29 & \ggx{+1.14} \\
Policy Change & 4.16 & 3.07 & \ggx{+1.09} & 2.84 & \ggx{+1.32} \\
Do Not Track & 0.24 & 0.27 & \rrx{-0.03} & 0.49 & \rrx{-0.25} \\
Specific Audiences & 5.47 & 8.44 & \rrx{-2.97} & 10.71 & \rrx{-5.24} \\
\xl{1}
Overall & 75.58 & 69.37 & \ggx{+6.21} & 71.01 & \ggx{+4.57}\\
\xl{2}
\end{tabular}}
\end{table}

The comparison shows that AR/VR applications generally have a higher percentage of highlighted words in categories like 3rd Party Sharing, Data Security, and Policy Change than free and premium websites. This reflects a focused effort to emphasize key privacy-related terms in these areas.

\BfPara{Groups Comparison}
In addition to the overall comparison, the highlighted words for different groups of AR/VR applications were also evaluated. Table \ref{tab:highlighted_words_groups} shows the percentage of highlighted words for each group.

\begin{table}[ht!]
\centering
\caption{Comparison of the mean distribution value (\boldmath$\mu$) of the highlighted words in \ours groups, free websites groups, and premium websites groups privacy policies.}
\label{tab:highlighted_words_groups}
\scalebox{0.99}{
\begin{tabular}{ lccc }
\xl{2}
{\bf Category} & {\boldmath$\mu$ \bf \ours} & {\boldmath$\mu$ \bf Free} & {\boldmath$\mu$ \bf Premium} \\
\xl{1}
1st Party Use & 98.28 & 89.69 & 94.64 \\
3rd Party Sharing & 85.05 & 86.69 & 89.00 \\
User Choice & 64.86 & 59.06 & 79.71 \\
User Access & 38.82 & 44.33 & 67.20 \\
Data Retention & 38.73 & 34.92 & 59.61 \\
Data Security & 69.58 & 67.97 & 75.78 \\
Policy Change & 75.19 & 72.41 & 71.75 \\
Do Not Track & 8.82 & 12.25 & 22.87 \\
Specific Audiences & 63.72 & 68.97 & 75.68 \\
\xl{2}
\end{tabular}}
\end{table}

The group comparison highlights that AR/VR applications emphasize certain privacy-related terms more consistently across different groups. Categories like 3rd Party Sharing and Data Security show higher percentages of highlighted words, indicating a strong emphasis on these aspects.

\BfPara{Observations and Explanations}
The focus on highlighted words in AR/VR applications can be attributed to several factors: 
\begin{enumerate*}
    \item \textbf{Key Term Emphasis}: Highlighting specific terms helps clearly communicate critical privacy practices to users.
    \item \textbf{Regulatory Requirements}: Emphasizing certain terms might be driven by regulatory guidelines that mandate clear communication of privacy practices.
    \item \textbf{User Trust}: Using highlighted words to emphasize key privacy aspects can enhance user trust and confidence in the application.
\end{enumerate*}

Overall, AR/VR applications demonstrate a deliberate effort to highlight key privacy terms, enhancing their privacy policies' clarity and effectiveness.

\section{Conclusion}\label{sec:conclusion}

This study emphasizes the importance of AR/VR developers creating transparent, user-friendly privacy policies. Developers can enhance privacy practices by utilizing advanced models like BERT and ensure clear communication of crucial information to users. The research findings reveal a shift in AR/VR privacy practices, focusing more on user trust and regulatory compliance. Practically, the results could guide the creation of better privacy policy tools for developers, addressing common issues highlighted in the study \cite{AlabduljabbarAMM21}. Additionally, these insights could help regulatory bodies establish stronger guidelines for AR/VR privacy standards \cite{BuiSCS21}.


\end{document}